\documentclass[12pt]{article}  
\usepackage{cite}
\usepackage{epsfig}
\usepackage{graphicx}
\usepackage{amsmath}
\usepackage{amssymb}
\usepackage{ulem}
\usepackage{mdwlist}          
\usepackage{color}              

\usepackage{a41}
\usepackage{color}
\usepackage[rflt]{floatflt}
\usepackage{float}
\usepackage{slashed}

\usepackage{array}

\usepackage[english]{babel}
\usepackage[latin1]{inputenc}
\usepackage[T1]{fontenc}
\usepackage{ae}

\usepackage{url}

\usepackage{amsmath, amsthm, amssymb}
\newtheorem{thm}{Theorem}[section]

\newtheorem{definition}[thm]{Definition}

 \newcommand{\GeV}{\mathrm{GeV}}
\newcommand{\Mvec}{{\rm\bf M}}
\newcommand{\MeV}{{\rm MeV}}

\newcommand{\ep}{\varepsilon}

\usepackage{rotating}

\usepackage{graphicx}
\newcounter{mmacnt}
\def\restartmma{\setcounter{mmacnt}{0}}
\restartmma \catcode`|=\active
\def|#1|{\mathrm{#1}}
\catcode`|=12
\newenvironment{mma}{
 \par\smallskip
 \catcode`|=\active
 \parskip=0pt\parindent=0pt % locally
 \small
 \def\In##1\\{%
   \def\linebreak{\hfill\break\null\qquad}%
   \refstepcounter{mmacnt}
   \hangindent=2.5em\hangafter=0
   \leavevmode
   \llap{\tiny\sffamily In[\arabic{mmacnt}]:=\kern.5em}%
   \mathversion{bold}\footnotesize$\displaystyle##1$\normalsize
   \mathversion{normal}\par
 }%
 \def\Print##1\\{%
   \def\linebreak{		\hfill\break}%
   \hangindent=2.5em\hangafter=0
   \leavevmode ##1\par}%
 \def\Out##1\\{%
   \def\linebreak{$\hfill\break\null\hfill$}%
   \kern\abovedisplayskip\par
   \hangindent=2.5em\hangafter=0
   \leavevmode
   \llap{\tiny\sffamily Out[\arabic{mmacnt}]=\kern.5em}
   \footnotesize$\displaystyle##1$\normalsize\hfill\null\par
   \kern\belowdisplayskip
 }%
 \def\Warning##1##2\\{%
   \def\linebreak{\hfill\break}%
   \hangindent=2.5em\hangafter=0
   \leavevmode
   {\scriptsize##1 : ##2}\par}%
}{%
 \par\smallskip
}

\usepackage{color}

\newenvironment{fshaded}{%
\MakeFramed {\FrameRestore}
}%
{\endMakeFramed}

\allowdisplaybreaks[4]

\begin{document}
\setlength{\baselineskip}{0.515cm}
\sloppy
\thispagestyle{empty}
\begin{flushleft}
DESY 21--096
\hfill 
%{\tt 	arXiv:2106.xxxxx [hep-ph]}
\\
DO--TH 21/21\\
SAGEX--21--13--E\\
June 2021\\
\end{flushleft}

\mbox{}
\vspace*{\fill}
\begin{center}

{\Large\bf The \boldmath N$^3$LO Scheme-invariant QCD Evolution of the Non-} 

\vspace*{3mm} 
{\Large\bf singlet Structure Functions \boldmath $F^{\rm NS}_2(x,Q^2)$ and \nolinebreak $g_1^{\rm 
NS}(x,Q^2)$}

\vspace{3cm}
\large
J.~Bl\"umlein and 
M.~Saragnese 

\vspace{1.cm}
\normalsize

{\it  Deutsches Elektronen--Synchrotron, DESY,}\\
{\it  Platanenallee 6, D-15738 Zeuthen, Germany}

%%\today

\end{center}
\normalsize
\vspace{\fill}
\begin{abstract}
\noindent
We present the scheme-invariant unpolarized and polarized flavor non--singlet evolution equation to 
N$^3$LO  for the structure functions $F_2(x,Q^2)$ and $g_1(x,Q^2)$ including the charm- and bottom quark
effects in the asymptotic representation. The corresponding evolution is based on the experimental measurement 
of the non--singlet structure functions at a starting scale $Q_0^2$. In this way the evolution does only depend 
on the strong coupling constant $\alpha_s(M_Z)$ or the QCD scale $\Lambda_{\rm QCD}$ and the charm and bottom quark 
masses $m_c$ and $m_b$ and provides one of the cleanest ways to measure the strong coupling constant in future 
high luminosity deep--inelastic scattering experiments. The yet unknown parts of the 4--loop anomalous dimensions 
introduce only a marginal error in this analysis.
\end{abstract}

\vspace*{\fill}
\noindent
%\numberwithin{equation}{section}

\newpage
%-----------------------------------------------------------------------------------------------------------
\section{Introduction}
\label{sec:1}
%-----------------------------------------------------------------------------------------------------------

\vspace*{1mm}
\noindent
The measurement of the strong coupling constant $\alpha_s(M_Z)$ from precision data on deep--inelastic scattering 
is one of the cleanest ways to obtain this fundamental parameter of the Standard Model. The present world data
allow measurements of $\delta \alpha_s(M_Z)/\alpha_s(M_Z)$ at the level of $\sim O(1\%)$, \cite{Bethke:2011tr,Alekhin:2016evh,
Moch:2014tta}. In the standard analyses one is fitting the scaling violations of deep--inelastic structure 
functions and obtains $\alpha_s(M_Z)$ together with the parameters of the parton distribution functions, observing 
the correlation of all parameters, cf. \cite{Accardi:2016ndt}. Compared to this, the measurement of $\alpha_s(M_Z)$ 
from $R_{e^+e^-}$, cf. \cite{Baikov:2012er}, does not need to fit a larger amount of further parameters.
The different precision determinations of $\alpha_s(M_Z)$ using different processes do not yet agree and further
precision measurements are therefore needed.

For deep--inelastic scattering a direct determination of $\alpha_s(M_Z)$ is also possible in the case the starting 
distributions of the evolution are measured experimentally. In the flavor non--singlet case the corresponding 
non--singlet structure functions at a scale $Q_0^2$ provide the input and the data at $Q^2 > Q_0^2$ are fitted
to a physical evolution function, which only depends on $\alpha_s(M_Z)$, provided that the input for the values 
of the heavy quark masses $m_c$ and $m_b$ is known at high precision, as also required in the measurement of 
$\alpha_s(M_Z)$ from $R_{e^+e^-}$. Here $Q^2$ denotes the virtuality of the exchanged gauge 
boson.
Therefore scheme--invariant evolution equations allow for a direct determination
of the strong coupling constant using measured one dimensional input distributions which are here functions of the
Bjorken variable $x$. An analysis of this kind has been performed in Ref.~\cite{Blumlein:2006be} in the 
unpolarized case, effectively up to N$^3$LO, however, only considering massless quarks with some further 
assumptions, see 
also Ref.~\cite{Blumlein:2012se}. The sensitivity of the structure function on  $\alpha_s(M_Z)$ or $\Lambda_{\rm 
QCD}$ is traditionally illustrated by the slope $\partial \ln(F_i(x,Q^2))/\partial \ln(Q^2)$ determined experimentally 
\cite{Eisele:1986uz,Benvenuti:1989rg,Blumlein:1989pd}.

In the present paper we complete the formalism by the single- and double--mass heavy flavor corrections for the 
Wilson coefficients up to three--loop order and provide numerical results for these effects, using the results 
given in Refs.~\cite{Buza:1995ie,Bierenbaum:2007qe,Buza:1996xr,Behring:2014eya,Ablinger:2014vwa,Behring:2015zaa,
Blumlein:2021xlc}\footnote{A numeric implementation of the heavy flavor Wilson coefficients to two--loop order 
has been given in \cite{Alekhin:2003ev}.}, as well as the massless contributions. 
Scheme--invariant evolution is usually performed in Mellin $N$ 
space for technical reasons, because it is simpler in analytic form compared to the corresponding $z$-space
representation. In the polarized case the reference to factorization scheme--invariant quantities will also 
solve the associated $\gamma_5$ problem, which usually arises performing the loop calculation in $D = 4 + \ep$
space--time dimensions. The strong coupling constant is used in the $\overline{\sf MS}$ scheme. A main goal of 
the present study is to outline the different evolution effects for high precision measurements. 

The measurement of scaling violations concerns that of massless parton evolution and is there implied by the 
ultraviolet singularity of the respective twist--2 local operators. Referring to the scale evolution using observables 
necessarily requests to map to the massless partons, which implies in $N$ space the algebraic use of the Wilson 
coefficients, already at the hard scale $Q^2$, as an obstacle, which are different for the various hard processes. For 
the latter ones it has to be guaranteed that only twist--2 terms contribute, 
cf.~\cite{Blumlein:2008kz,Alekhin:2012ig,Blumlein:2012se}, by putting the cuts $W^2 > 15~\GeV^2, Q^2 >  10~\GeV^2$, 
where $W$ is the 
hadronic mass. The natural scale of evolution in  deep--inelastic scattering is $Q^2$.

The evolution equation is solved analytically and one final numerical contour integral around the 
singularities of the corresponding problem delivers the evolved flavor non--singlet structure function in 
$x$ space. We first consider the flavor decomposition to provide realistic input distributions and derive
the corresponding evolution operators for the structure function. Here we also discuss the remainder systematics
of the yet not completely known four--loop non--singlet splitting functions $P^{\rm NS, (3), \pm}$. 
Finally we are providing numerical results in the unpolarized and polarized case. Precision analyses of this
kind can be performed in the physics programme at the future collider experiments at the EIC \cite{Boer:2011fh} 
and  LHeC \cite{AbelleiraFernandez:2012cc}.
%-----------------------------------------------------------------------------------------------------------
\section{Flavor Decomposition}
\label{sec:2}
%-----------------------------------------------------------------------------------------------------------

\vspace*{1mm}
\noindent
Since the evolution for the three different flavor non--singlet distributions and the singlet evolution are 
different,
the experimental input at the starting scale $Q_0^2$ has to be projected correspondingly by combining different 
deep--inelastic structure functions. One may refer to the following well--known decomposition, see also 
\cite{Vogt:2004ns}. Let
%-----------------------------------------------------------------------------------------------------------
\begin{eqnarray}
v_{k^2-1}^{\pm} &=& \sum_{l=1}^k (q_l \pm \bar{q}_l) - k (q_k \pm \bar{q}_k),
\end{eqnarray}
%-----------------------------------------------------------------------------------------------------------
with $q_i$ the quark distributions and
%-----------------------------------------------------------------------------------------------------------
\begin{eqnarray}
v_{1}^{\pm} &=& 0
\\
v_{3}^{\pm} &=& (u \pm \bar{u}) - (d \pm \bar{d}) 
\\
v_{8}^{\pm} &=& (u \pm \bar{u}) + (d \pm \bar{d}) - 2(s \pm \bar{s}),
\end{eqnarray}
%-----------------------------------------------------------------------------------------------------------
one has
%-----------------------------------------------------------------------------------------------------------
\begin{eqnarray}
q_i + \bar{q}_i &=& \frac{1}{N_F} \Sigma - \frac{1}{i} v^+_{i^2-1} + \sum_{l=i+1}^{N_F} \frac{1}{l(l-1)} 
v^+_{l^2-1},
\\
\Sigma &=& \sum_{l=1}^{N_F} (q_l + \bar{q}_l).
\end{eqnarray}
%-----------------------------------------------------------------------------------------------------------
The nucleon structure functions for pure photon exchange in the case of three light flavors $(u, d, s)$ are then 
given at leading order (LO) by
%-----------------------------------------------------------------------------------------------------------
\begin{eqnarray}
F_2^p  &=& x\left[\frac{2}{9} \Sigma + \frac{1}{6} v_3^+  + \frac{1}{18} v_8^+\right]
\\
F_2^d  &=& \frac{1}{2}\left[F_2^p + F_2^n\right] = x \left[\frac{2}{9} \Sigma +   \frac{1}{18} 
v_8^+\right].
\end{eqnarray}
%-----------------------------------------------------------------------------------------------------------
A synonymous decomposition applies to the structure functions $xg_1^p$ and $xg_1^d$. To project onto the singlet 
distribution $\Sigma$ 
directly one usually needs
charged current structure functions, as measured in neutrino experiments \cite{Boscolo:2018ytm} 
or at facilities
like the planned LHeC project \cite{AbelleiraFernandez:2012cc} or the EIC \cite{Boer:2011fh} in the 
unpolarized and polarized case, at high luminosity, with~\cite{Blumlein:2012bf}
%-----------------------------------------------------------------------------------------------------------
\begin{eqnarray}
\frac{1}{2}\left[W_2^{p,+} + W_2^{p,-}\right]  &=& x \Sigma
\end{eqnarray}
%----------------------------------------------------------------------------------------------------------
above the charm threshold \cite{SCHMITZ}. Here the index $\pm$ denotes the exchange of a $W^+$ or a $W^-$ boson, 
respectively. Otherwise additional information on sea--quarks is necessary. The flavor 
non--singlet combinations we 
are considering are given by
%-----------------------------------------------------------------------------------------------------------
\begin{alignat}{2}
\label{F2:decomp}
F_2^{\rm NS} &= F_2^p  - F_2^d   &&= \frac{1}{6} x C_q^{\rm NS,+} \otimes v_3^+
\\
\label{g1:decomp}
x g_1^{\rm NS} &= 
x g_1^p  - x g_1^d  
&&= \frac{1}{6} x \Delta C_q^{\rm NS,+} \otimes \Delta v_3^+,
\end{alignat}
%-----------------------------------------------------------------------------------------------------------
where the Wilson coefficients are $C_q$ and $\Delta C_q$ contain light and heavy flavor contributions,  and
$\otimes$ denotes the Mellin convolution
%-----------------------------------------------------------------------------------------------------------
\begin{eqnarray}
\label{eq:MC}
[A \otimes B](x) = \int_0^1 dx_1 \int_0^1 dx_2 \delta(x -x_1 x_2) A(x_1) B(x_2).
\end{eqnarray}
%-----------------------------------------------------------------------------------------------------------
The Mellin transform
%-----------------------------------------------------------------------------------------------------------
\begin{eqnarray}
\Mvec[f(x)](N) = \int_0^1 dx x^{N-1} f(x)
\end{eqnarray}
%-----------------------------------------------------------------------------------------------------------
turns Eq.~(\ref{eq:MC}) into the product
%-----------------------------------------------------------------------------------------------------------
\begin{eqnarray}
\Mvec[(A \otimes B)(x)](N) = \Mvec[A(x))](N) \cdot \Mvec[B(x))](N).
\end{eqnarray}
%-----------------------------------------------------------------------------------------------------------
Before forming the structure function difference in (\ref{F2:decomp}, \ref{g1:decomp}) one has to unfold the
nuclear corrections of the deuteron structure functions. The lowest Mellin moment of (\ref{g1:decomp})
is given by the polarized Bjorken sum rule \cite{Bjorken:1969mm}.
%-----------------------------------------------------------------------------------------------------------
\section{The Non--singlet Evolution}
\label{sec:3}
%-----------------------------------------------------------------------------------------------------------

\vspace*{1mm}
\noindent
The evolution operator of scheme--invariant flavor non--singlet evolution, $E_{\rm NS}$, is obtained as 
follows.\footnote{Here and below we will work in Mellin $N$ space.}

The evolution equation for the non--singlet structure functions can be written as
%-----------------------------------------------------------------------------------------------------------
\begin{eqnarray}
\label{eq:FNS1}
\frac{d}{d \ln(Q^2)} \ln\left[F^{\rm NS}(Q^2)\right] &=& 
  \frac{d}{d \ln(Q^2)} \ln\left[C^{\rm NS}(Q^2)\right] 
+ \frac{d}{d \ln(Q^2)} \ln\left[q^{\rm NS}(Q^2)\right].
\end{eqnarray}
%-----------------------------------------------------------------------------------------------------------
Its solution is given by
%-----------------------------------------------------------------------------------------------------------
\begin{eqnarray}
\label{eq:FNS2}
F^{\rm NS}(Q^2) &=& E_{\rm NS} (Q^2,Q_0^2) \cdot F^{\rm NS}(Q^2_0).
\end{eqnarray}
%-----------------------------------------------------------------------------------------------------------
The Wilson coefficient is given by
%-----------------------------------------------------------------------------------------------------------
\begin{eqnarray}
C(Q^2) &=& 1 + \sum_{k=1}^\infty a_s^k(Q^2) C_k,~~~~~C_k = c_k   + h_k\left(L_c,L_b\right). 
\end{eqnarray}
%-----------------------------------------------------------------------------------------------------------
Here $c_k$ denote the expansion coefficients of the massless Wilson coefficients and $h_k$ of the massive Wilson 
coefficient, with 
%-----------------------------------------------------------------------------------------------------------
\begin{eqnarray}
L_c = \ln\left(\frac{Q^2}{m_c^2}\right),~~~~~~
L_b = \ln\left(\frac{Q^2}{m_b^2}\right)
\end{eqnarray}
%-----------------------------------------------------------------------------------------------------------
and $m_{c,b}$ are the on--shell charm and bottom quark masses.

In the non--singlet case the heavy flavor corrections contribute form
$O(a_s^2)$ onward. One has
%-----------------------------------------------------------------------------------------------------------
\begin{eqnarray}
h_1 &=& 0
\\
h_2 &=& \hat{h}_2(L_c) + \hat{h}_2(L_b)
\\
h_3 &=& \hat{h}_3(L_c) + \hat{h}_3(L_b) + \hat{\hat{h}}_3(L_c,L_b),
\end{eqnarray}
%-----------------------------------------------------------------------------------------------------------
where $\hat{h}_i$ denote the single mass and $\hat{\hat{h}}_3$ the double mass contributions. 

One may rewrite the differential operator
%-----------------------------------------------------------------------------------------------------------
\begin{eqnarray}
\label{eq:DIO}
\frac{d}{d \ln(Q^2)} =  \frac{d a_s(Q^2)}{d \ln(Q^2)} \cdot \frac{d}{d a_s(Q^2)} 
\end{eqnarray}
%-----------------------------------------------------------------------------------------------------------
with
%-----------------------------------------------------------------------------------------------------------
\begin{eqnarray}
\label{eq:AS}
\frac{d a_s}{d \ln(Q^2)} = - \sum_{k=0}^\infty \beta_k a_s^{k+2}.
\end{eqnarray}
%-----------------------------------------------------------------------------------------------------------

The evolution equation for the non--singlet quark density is given by
%-----------------------------------------------------------------------------------------------------------
\begin{eqnarray}
\frac{d}{d a_s} \ln\left[q^{\rm NS}(Q^2)\right] = - \frac{1}{2} \frac{\sum_{k=0}^\infty 
a_s^{k+1} P_k}{\sum_{k=0}^\infty \beta_k 
a_s^{k+2}},
\end{eqnarray}
%-----------------------------------------------------------------------------------------------------------
where $\beta_k$ are expansion coefficients of the QCD--$\beta$ function and $P_k$ are the splitting functions.
The anomalous dimensions are related to the splitting functions\footnote{Our normalizations are such that a factor 
of two has to be applied to those given in \cite{Moch:2004pa,Vogt:2004mw}.} 
by
%-----------------------------------------------------------------------------------------------------------
\begin{eqnarray}
\gamma_{ij}^{(k)}(N) = - \int_0^1 dx x^{N-1} P_{ij}^{(k)}(x). 
\end{eqnarray}
%-----------------------------------------------------------------------------------------------------------

The solution of Eq.~(\ref{eq:FNS1}) to N$^3$LO reads
%-----------------------------------------------------------------
\begin{eqnarray}
E_{\rm NS}(Q^2,Q_0^2) & =&\left(\frac{a}{ a_0 }\right)^{-\frac{ P_0 }{2\beta_0 }}
\Biggl\{
1
+\frac{a- a_0 }{2\beta_0 ^2} \biggl\{
\Bigl[1 + a^2  C_2(Q^2) - a_0 ^2  C_2(Q_0^2) \Bigr] \bigl(2 \beta_0 ^2  C_1 -\beta_0   P_1 +\beta_1  P_0 \bigr)
\nonumber\\&&    
-\frac{\bigl(a^2- a_0 ^2\bigr)}{4 \beta_0 ^3} \bigl(2\beta_0 ^2  C_1 -\beta_0   P_1 +\beta_1   P_0 \bigr) \Bigl[2\beta_0 ^3 C_1 ^2+\beta_0 ^2 P_2 - \beta_0 \beta_1 P_1 + \bigl(\beta_1 ^2-\beta_0  \beta_2 \bigr) P_0 \Bigr]
\nonumber\\&&
+\frac{\bigl(a^2+a  a_0 + a_0 ^2\bigr)}{3 \beta_0 ^2} \Bigl[2\beta_0 ^4  C_1 ^3-\beta_0 ^3  P_3 +\beta_0 ^2 \beta_1 P_2 + \bigl(\beta_0 ^2 \beta_2 -\beta_0  \beta_1 ^2\bigr) P_1 
\nonumber\\&&
+ \bigl(\beta_0 ^2 \beta_3 -2 \beta_0  \beta_1  \beta_2 +\beta_1 ^3\bigr) P_0 \Bigr]
+\frac{a- a_0 }{4 \beta_0 ^2} \bigl(2\beta_0 ^2  C_1 -\beta_0   P_1 +\beta_1   P_0 \bigr)^2
\nonumber\\&&
+\frac{(a- a_0 )^2}{24 \beta_0 ^4} \bigl(2\beta_0 ^2  C_1 -\beta_0   P_1 +\beta_1   P_0 \bigr)^3
-\frac{a+ a_0 }{2 \beta_0 } \Bigl[2\beta_0 ^3  C_1 ^2+\beta_0 ^2  P_2 -\beta_0  \beta_1 P_1 
\nonumber\\&&
+ P_0  \bigl(\beta_1 ^2-\beta_0  \beta_2 \bigr)
 \Bigr]
\biggr\}
+a^2  C_2(Q^2) - a_0 ^2  C_2(Q_0^2) 
- C_1  \Bigl[a^3  C_2(Q^2) - a_0 ^3  C_2(Q_0^2) \Bigr]
\nonumber\\&&
+ a^3 C_3(Q^2) - a_0 ^3  C_3(Q_0^2) 
\Biggr\}
\label{eq:ENS}
\end{eqnarray}
%-----------------------------------------------------------------
%-----------------------------------------------------------------------------------------------------------
and $a = a_s(Q^2),~a_0 = a_s(Q^2_0)$ and $P_i(N) \equiv P_i$ denotes the Mellin transform of $P_i(z)$.
The heavy quark contributions to 
the  Wilson coefficients 
are given by \cite{Blumlein:2016xcy,Ablinger:2014vwa,Behring:2015zaa,Bierenbaum:2009mv}
%-----------------------------------------------------------------------------------------------------------
\begin{eqnarray}
\label{H2}
\hat{h}_2^{(Q)} &=& - \frac{\beta_{0,Q}}{4} P_{qq}^{(0)} \ln^2\left(\frac{Q^2}{m^2}\right)
                   + \frac{1}{2} \hat{P}_{qq}^{(1), \rm NS} \ln\left(\frac{Q^2}{m^2}\right)
                   + a_{qq}^{(2), \rm NS} + \frac{\beta_{0,Q}}{4} \zeta_2 P_{qq}^{(0)} + \hat{C_q}^{(2),\rm NS}
\\
%------------------------
\label{H3}
\hat{h}_3^{(Q)} &=& -\frac{1}{6} P_{qq}^{(0)} \beta_{0,Q} \left(\beta_0 + 2 \beta_{0,Q}\right) 
\ln^3\left(\frac{Q^2}{m^2}\right)
+ \frac{1}{4} \Biggl[-2 P_{qq}^{(1), \rm NS} \beta_{0,Q} + 2 \hat{P}_{qq}^{(1), \rm NS} \left(\beta_0 + 
\beta_{0,Q}\right) 
\nonumber\\ && 
- \beta_{1,Q} P_{qq}^{(0)} \Biggr] \ln^2\left(\frac{Q^2}{m^2}\right)
- \frac{1}{2} \Biggl[- \hat{P}_{qq}^{(2),\rm NS} - \left(4 a_{qq,Q}^{(2),\rm NS} + \zeta_2 \beta_{0,Q} 
P_{qq}^{(0)}\right)  \left(\beta_0 +  \beta_{0,Q}\right) 
\nonumber\\ && 
- P_{qq}^{(0)} \beta_{1,Q}^{(1)}\Biggr] 
\ln\left(\frac{Q^2}{m^2}\right)
+ 4 \bar{a}_{qq,Q}^{(2),\rm NS}(\beta_0+\beta_{0,Q}) + P_{qq}^{(0)} \beta_{1,Q}^{(2)} 
+ \frac{1}{6} P_{qq}^{(0)} \beta_0 \beta_{0,Q} \zeta_3 
\nonumber\\ && 
+ \frac{1}{4} P_{qq}^{(1),\rm NS} \beta_{0,Q} \zeta_2 
- 2 \delta m_1^{(1)} \beta_{0,Q} P_{qq}^{(0)} - \delta m_1^{(0)} \hat{P}_{qq}^{(1),\rm NS} + 2 \delta m_1^{(-1)} 
a_{qq,Q}^{(2),\rm NS} + a_{qq,Q}^{(3),\rm NS} 
\nonumber\\ && + \Biggl[- \frac{\beta_{0,Q}}{4} P_{qq}^{(0)} \ln^2\left(\frac{Q^2}{m^2}\right)
                   + \frac{1}{2} \hat{P}_{qq}^{(1), \rm NS} \ln\left(\frac{Q^2}{m^2}\right)
                   + a_{qq}^{(2), \rm NS} + \frac{\beta_{0,Q}}{4} \zeta_2 P_{qq}^{(0)}\Biggr] C_q^{(1),\rm NS}
\nonumber\\ &&
                   + \hat{C}_q^{(3),\rm NS}.  
\end{eqnarray}
%-----------------------------------------------------------------------
The two--mass three--loop contributions \cite{Ablinger:2017err} read 
%-----------------------------------------------------------------------
\begin{eqnarray}
\label{H23}
\hat{\hat{h}}_3^{\rm NS} &=& P_{qq}^{(0)} \beta_{0,Q}^2\left[\frac{2}{3} \left(L_c^3 + L_b^3\right) + \frac{1}{2} 
\left(L_c^2 L_b +L_c L_b^2 \right)\right] - \beta_{0,Q} \hat{P}_{qq}^{(1),\rm NS} \left(L_c^2 + L_b^2\right)
-\left[4 a_{qq,Q}^{(2),\rm NS} \beta_{0,Q} \right. 
\nonumber\\ && \left.
- \frac{1}{2} \beta_{0,Q}^2 P_{qq}^{(0)} \zeta_2\right] (L_c + L_b)
+ 8 \bar{a}_{qq,Q}^{(2),\rm NS} \beta_{0,Q} + \tilde{a}^{(3),\rm NS}_{qq,Q}(m_c,m_b,Q^2).
\end{eqnarray}
%-----------------------------------------------------------------------
The two--mass term is the same in the unpolarized and polarized case.
In the r.h.s. of Eqs.~(\ref{H2}--\ref{H23}) we define
%-----------------------------------------------------------------------------------------------------------
\begin{eqnarray}
\hat{f}(x,N_F) = f(x,N_F+1) - f(x,N_F).
\end{eqnarray}
%-----------------------------------------------------------------------

\noindent
In the $\overline{\sf MS}$ scheme the iterative solution for $a_s(Q^2)$ is \cite{Chetyrkin:1997sg} 
%-----------------------------------------------------------------------------------------------------------
\begin{eqnarray}
a_s(Q^2) &=& \frac{1}{\beta_0 L} 
- \frac{\beta_1}{\beta_0^3 L^2} \ln(L)
+ \frac{1}{\beta_0^3 L^3} \Biggl[\frac{\beta_1^2}{\beta_0^2 }(\ln^2(L) - \ln(L) -1) + \frac{\beta_2}{\beta_0}\Biggr]
\nonumber\\ &&
+ \frac{1}{\beta_0^4 L^4} \Biggl[\frac{\beta_1^3}{\beta_0^3 }\Biggl(-\ln^3(L) + \frac{5}{2} \ln^2(L) + 2 \ln(L) - 
\frac{1}{2}\Biggr) - 3 \frac{\beta_1 \beta_2}{\beta_0^2} \ln(L) + \frac{\beta_3}{2 \beta_0}\Biggr],
\end{eqnarray}
%-----------------------------------------------------------------------------------------------------------
with $L = \ln(Q^2/\Lambda_{\rm QCD}^2)$.  Here the integration constant for solving (\ref{eq:AS}) is chosen by 
$(\beta_1/\beta_0^2) \ln(\beta_0)$ \cite{Bardeen:1978yd}. The expansion coefficients of the 
$\beta$-function to N$^3$LO were calculated in \cite{BETA}.
The flavor matching conditions were given in \cite{Chetyrkin:1997sg}. 
The expansion coefficients of the renormalized mass were given in  \cite{Gray:1990yh,Broadhurst:1991fy}.
The constant and $O(\ep)$ parts of the massive unrenormalized operator matrix elements at $O(a_s^k)$
are denoted by $a_{ij}^{(k)}$ and $\bar{a}_{ij}^{(k)}$, respectively, cf.~\cite{
Buza:1995ie,Bierenbaum:2007qe,Bierenbaum:2008yu,POL21}.

The three--loop massless Wilson coefficients are expressed by effective representations. Otherwise we use the analytic 
Mellin space representations. After algebraic reduction \cite{Blumlein:2003gb},
they depend on 32 harmonic sums \cite{Vermaseren:1998uu,Blumlein:1998if} up to weight 
{\sf w=6} only,  which is particular to the flavor non--singlet case 
to three--loop orders. Other massive Wilson coefficients have a more involved structure 
\cite{Ablinger:2014nga,Ablinger:2014uka,AGG,Ablinger:2017ptf}. It is useful to represent 
harmonic sums 
with an alternating index by their Mellin transform, to eliminate spurious $(-1)^N$ terms \cite{Blumlein:1998if} before 
performing the analytic continuation from even or odd integers to the complex plane. The singularities of the problem 
are located at the non--positive integers, around which the contour integral is performed, 
cf.~\cite{Diemoz:1987xu,Gluck:1989ze,Blumlein:1997em,Ellis:1993rb}. For the analytic continuation we follow 
Refs.~\cite{Blumlein:2009fz,Blumlein:2009ta}.

In the case of the structure function $F_2$ the non--singlet anomalous dimensions are $P^{{\rm NS},+}$ and the 
expansion coefficients of the Wilson coefficient up to $c_3$ were calculated in \cite{Vermaseren:2005qc}.
In the case of the structure function $g_1$ the anomalous dimensions are $P^{{\rm NS},-}$, cf. \cite{Moch:2004pa}, 
and the Wilson coefficient has been given in \cite{Moch:2008fj}.\footnote{Lower orders of the non--singlet anomalous 
dimensions and 
the massless non--singlet Wilson coefficients have been calculated in 
\cite{Gross:1973ju,
Georgi:1951sr,
Floratos:1977au,
GonzalezArroyo:1979df,
GonzalezArroyo:1979ng,
Curci:1980uw,
Floratos:1980hk,
Floratos:1981hs,
Moch:1999eb,
Ablinger:2014vwa,Behring:2019tus}
and 
\cite{Furmanski:1981cw,
vanNeerven:1991nn,
Zijlstra:1992kj,
Zijlstra:1992qd,
Moch:1999eb
} and in the polarized 
case in \cite{Sasaki:1975hk,
Ahmed:1975tj,
Altarelli:1977zs,
Mertig:1995ny,
Vogelsang:1995vh,
Vogelsang:1996im} and 
\cite{Kodaira:1978sh,
Kodaira:1979ib,
Antoniadis:1980dg,
Zijlstra:1993sh}.}
Below we will also use the combination
%-----------------------------------------------------------------------------------------------------------
\begin{eqnarray}
F_2^{\rm h}(N,Q^2) = \left[E_{\rm NS} - \left. E_{\rm NS}\right|_{\rm h=0}\right] F_2(N,Q_0^2).
\end{eqnarray}
%-----------------------------------------------------------------------------------------------------------

The four-loop non--singlet anomalous dimensions are not yet completely known as a function of $N$. However,
a series of moments has been calculated in \cite{Baikov:2006ai,Baikov:2015tea,Velizhanin:2014fua,Ruijl:2016pkm,
Davies:2016jie,Moch:2017uml}. We follow the earlier investigation in Ref.~\cite{Blumlein:2006be} and compare these 
moments with the Pad\'e-approximation
%-----------------------------------------------------------------------------------------------------------
\begin{eqnarray}
\label{eq:PADE}
P_{qq}^{3, \pm, \rm NS}(N) \approx \frac{P_{qq}^{2, \pm, \rm NS}(N)^2}{P_{qq}^{1, \pm, \rm NS}(N)},
\end{eqnarray}
%-----------------------------------------------------------------------------------------------------------
with the exact moments in 
\cite{Baikov:2006ai,Baikov:2015tea,Velizhanin:2014fua,Ruijl:2016pkm,Davies:2016jie,Moch:2017uml} 
in Table~\ref{TAB1}.  
Furthermore, the leading $N_F$ terms for the even moments have been predicted in \cite{Gracey:1994nn}.
 
From the 2nd moment, which agrees within $21 \%$, cf. also \cite{Blumlein:2006be}, the accuracy improves to 2.2 \% 
for the known even moments at $N=16$ and to 2.6\% for the odd moments at $N=15$. For the first moment for 
$\gamma^{-,\rm NS}$ the Pad\'e-approximation delivers even the correct result, after using the l'Hospital rule.
%-----------------------------------------------------------------------------------------------------------
\begin{table}[H]\centering
\renewcommand{\arraystretch}{1.3}
\begin{tabular}{|r|l|r|l|}
\hline
\multicolumn{1}{|c}{$N$} & 
\multicolumn{1}{|c}{$\delta \gamma^{+,\rm NS}$} & 
\multicolumn{1}{|c}{$N$} & 
\multicolumn{1}{|c|}{$\delta \gamma^{-,\rm NS}$} \\
\hline
 2 & 0.208822541 &  1 &  0.0          \\
 4 & 0.123728742 &  3 &  0.147102092  \\
 6 & 0.087155544 &  5 &  0.101634935  \\
 8 & 0.064949195 &  7 &  0.074593595  \\
10 & 0.049680399 &  9 &  0.056598595  \\
12 & 0.038394815 & 11 &  0.043633919  \\
14 & 0.029638565 & 13 &  0.033767853  \\
16 & 0.022602035 & 15 &  0.025956941  \\
\hline
\end{tabular}
\caption[]{
\label{TAB1}
\sf The relative error comparing the exact moments of the four--loop anomalous dimensions, $\gamma^{(3),\pm, \rm 
NS}$, with the Pad\'e approximation (\ref{eq:PADE}).}
\renewcommand{\arraystretch}{1.0}
\end{table}
%-----------------------------------------------------------------------------------------------------------
The leading small $x$ terms for $P_3^{\rm NS,+}$ and $P_2^{\rm NS,-}$ have been given in
\cite{Kirschner:1983di} after correcting some misprints there in Ref.~\cite{Blumlein:1995jp}, see also
\cite{Bartels:1995iu}. However, yet unknown sub--leading terms, having been studied at the known lower orders, 
are numerically dominant over the leading small $x$ corrections and  several of these sub--leading terms are needed, 
cf.~\cite{Blumlein:1995jp,Blumlein:1997em}, to be calculated to obtain reliable quantitative prediction in this
region of $x$.
Table~\ref{TAB1} shows that (\ref{eq:PADE}) provides an excellent model.
The earlier assumption of an error of 100\%  in \cite{Blumlein:2006be} on this relation has been very conservative and still
led to an error in $\Lambda_{\rm QCD}$ of  $2~\MeV$ only, which could now be improved in principle.
Yet the experimental
accuracy at this level cannot be reached at present, since the current experimental error amounts to 
$\delta \Lambda_{\rm QCD} = 26~\MeV$ \cite{Blumlein:2006be}. 
%-----------------------------------------------------------------------------------------------------------
\section{Numerical Results}
\label{sec:4}
%-----------------------------------------------------------------------------------------------------------

\vspace*{1mm}
\noindent
The measured input distributions contain correlated errors, which are parameterized by the 
respective fits. cf.~ 
\cite{Blumlein:2006be,Blumlein:2010rn,Alekhin:2012ig}. Their evolution, including the corresponding correlation matrix, 
has to be performed to provide 
the error prediction of the structure function at every fixed value of $\Lambda_{\rm QCD}, m_c$ and $m_b$ 
respectively. Here one may use the corresponding formulae for the correlated error propagation given in 
Ref.~\cite{Blumlein:2006be} and extensions of them, which are straightforward. Within future data analyses one 
will probably import the values of the heavy quark 
masses from the world data analyses. For the charm quark mass it has already been shown that its value obtained
from deep--inelastic scattering data fully agrees with other precision measurements \cite{Alekhin:2012vu}.
In our illustrations we will use the values $m_c = 1.59~\GeV$ \cite{Alekhin:2012vu} and $m_b = 4.78~\GeV$ \cite{PDG}.

For the input distribution in the unpolarized case we refer to the one of Ref.~\cite{Blumlein:2006be}
%-----------------------------------------------------------------------------------------------------------
\begin{eqnarray}
F_2^{\rm NS}(x, Q_0^2)  &=& C^{\rm NS}(x, Q_0^2) \otimes x q^{\rm NS}(x,Q_0^2)
\\
xq^{\rm NS}(x,Q_0^2) &=& 
\frac{1}{3} \biggl[
        0.262 \ x^{0.298} (1-x)^{4.032} (1+6.042 \sqrt{x} +35.49 x)
\nonumber\\&&        
           ~~~~~-1.085 \ x^{0.5} (1-x)^{5.921} (1-3.618 \sqrt{x} +16.41 x) \biggr] 
\end{eqnarray} 
%-----------------------------------------------------------------------------------------------------------
at $Q_0^2 = 4~\GeV^2$. In the polarized case we use the fit to $x g_1^{\rm NS}(x,Q_0^2)$ of the structure 
function
of Ref.~\cite{Blumlein:2010rn} at $Q_0^2 = 10~\GeV^2$,
%-----------------------------------------------------------------------------------------------------------
\begin{eqnarray}
x g_1^{\rm NS}(x,Q_0^2) &=& 2.4312 \cdot 10^{-5} x^{-0.38573} (1-x)^{2.69522} 
\nonumber\\ && \times (-1+279.624 \sqrt{x} + 1239.76 x + 7053.93 
x^{3/2} + 1866.208 x^2)
\end{eqnarray}
%-----------------------------------------------------------------------------------------------------------
%----------------------------------------------------------------------------
\begin{figure}[H]
        \centering
        \includegraphics[width=0.49\textwidth]{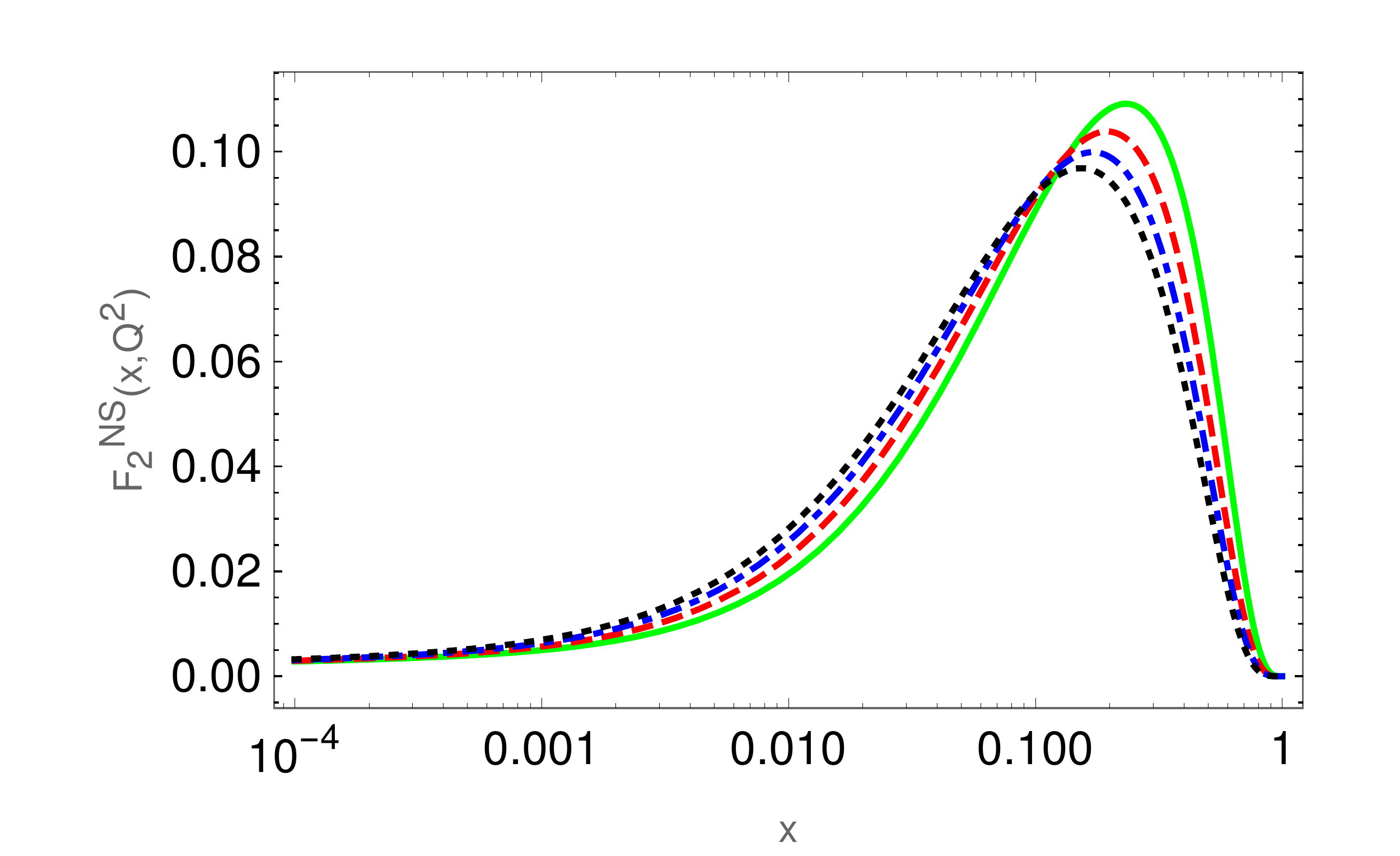}
        \includegraphics[width=0.49\textwidth]{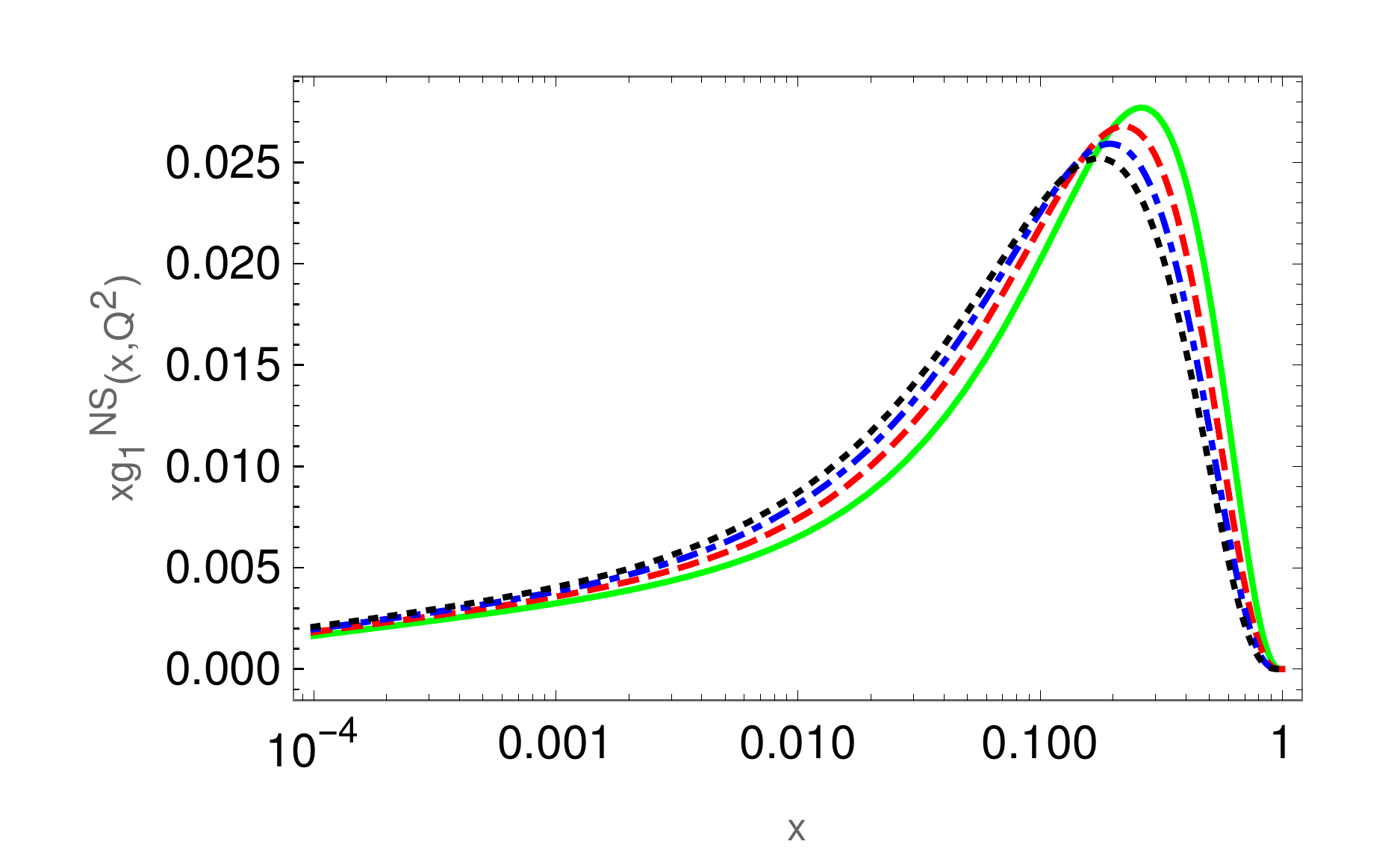}
        \caption{\sf Left:~The structure function $F_2^{\rm NS}$ at N$^3$LO. Right:~The structure function 
$xg_1^{\rm NS}$ at N$^3$LO. Full lines: $Q^2 = 10~\GeV^2$; dashed lines: $100~\GeV^2$;
dash-dotted lines: $1000~\GeV^2$; 
dotted lines: $10000~\GeV^2$.}
\label{fig1}
\end{figure}
%----------------------------------------------------------------------------
\noindent
In the numeric illustration our reference starting scale will be chosen to be $Q_0^2 = 10~\GeV^2$ both in the
unpolarized and polarized case.
In Figures~\ref{fig1} we show the scheme--invariant evolution of the non--singlet structure functions 
$F_2^{\rm NS}$ and $x g_1^{\rm NS}$ in the kinematic region $Q^2 \in [10, 10^4]~\GeV^2$. In Figures~\ref{fig2}
we expand the representation for the region of larger values of $x$. Both structure functions show a falling 
behaviour both towards small and large values of $x$. To improve the measurement of the strong coupling constant
future measurements of these structure functions are necessary at the 1\% level. 

%----------------------------------------------------------------------------
\begin{figure}[H]
        \centering
        \includegraphics[width=0.49\textwidth]{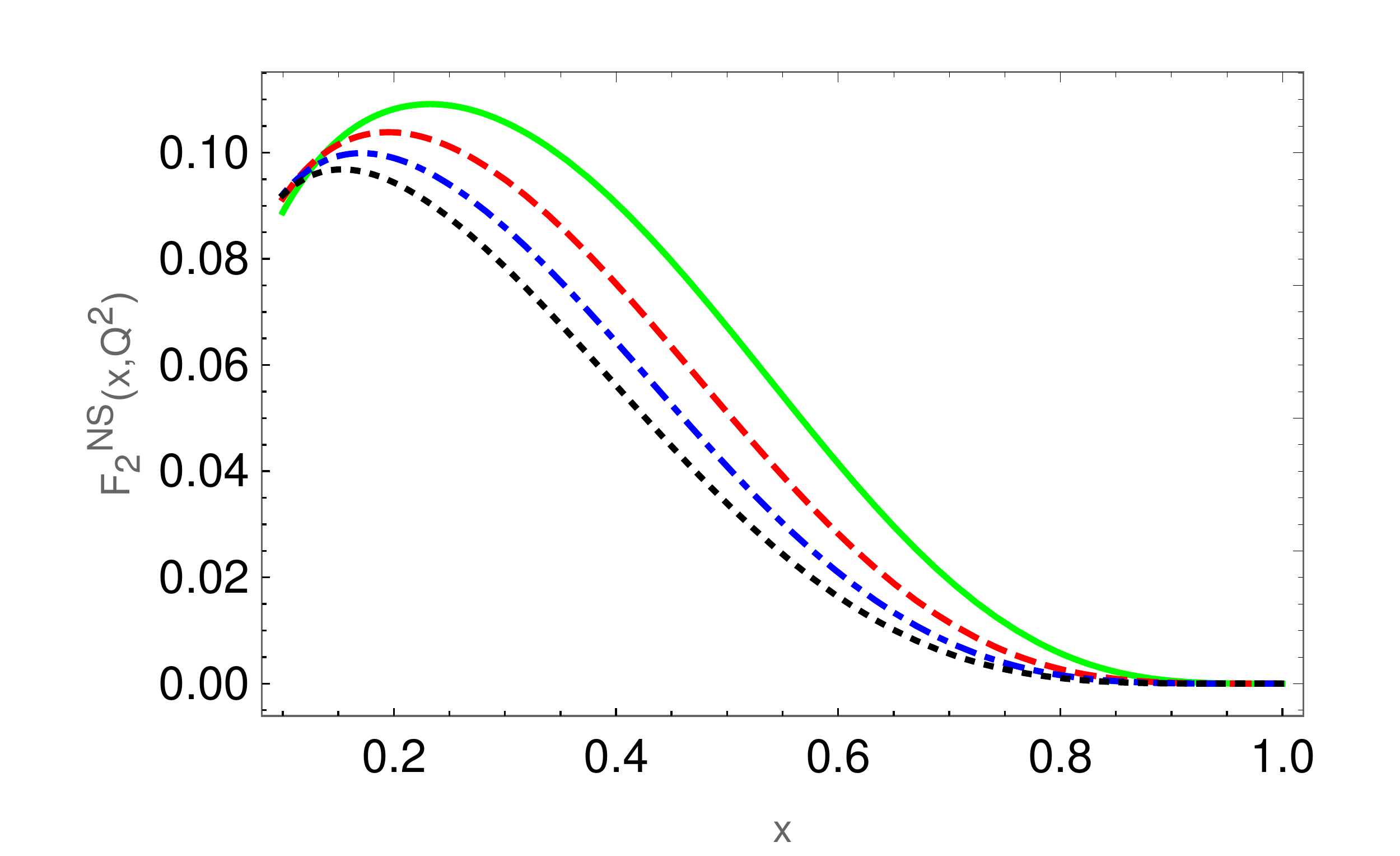}
        \includegraphics[width=0.49\textwidth]{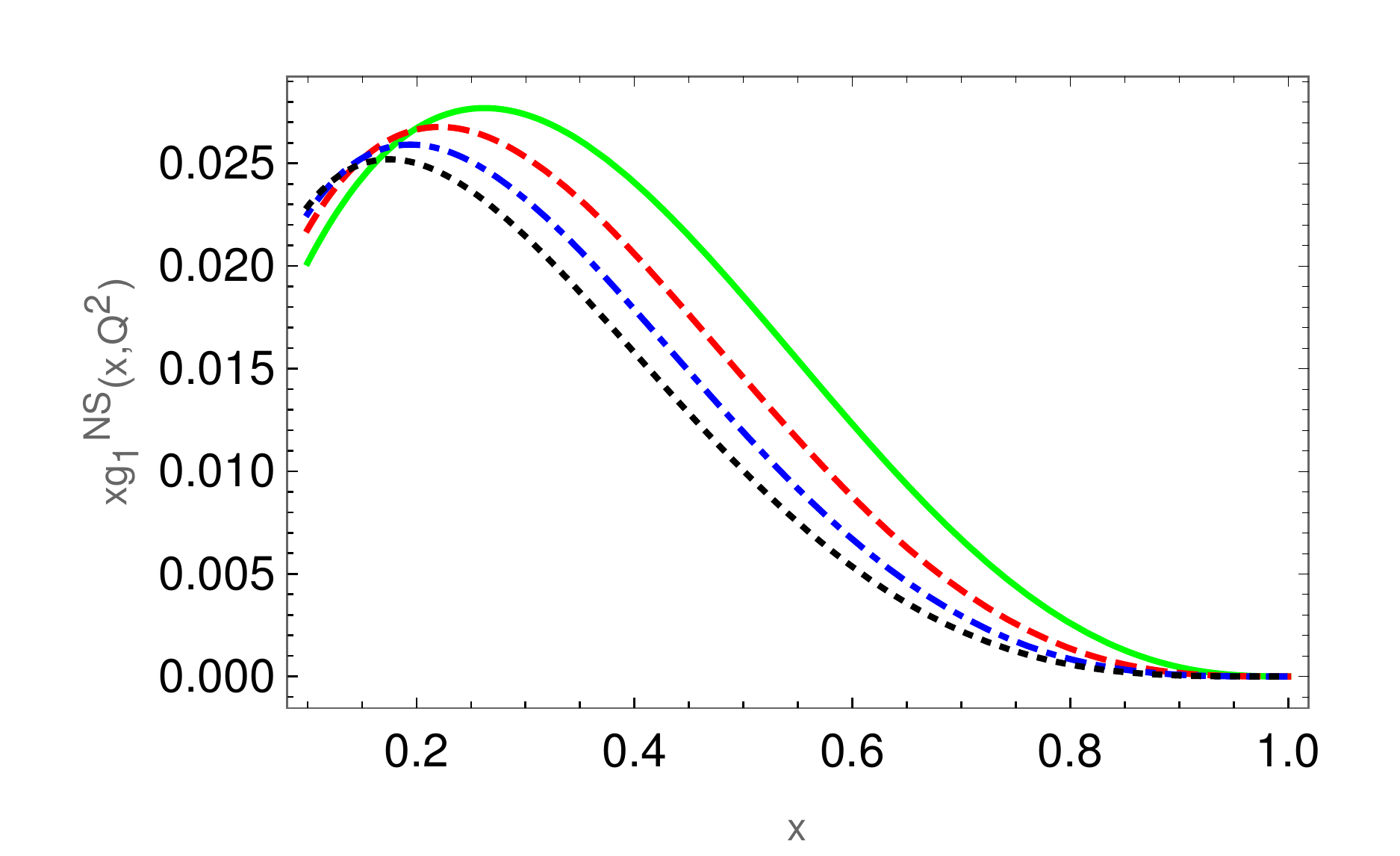}
        \caption{\sf The same as Figure~1 but with expanded large $x$ region.}
\label{fig2}
\end{figure}
%----------------------------------------------------------------------------

%----------------------------------------------------------------------------
\begin{figure}[H]
        \centering
        \includegraphics[width=0.49\textwidth]{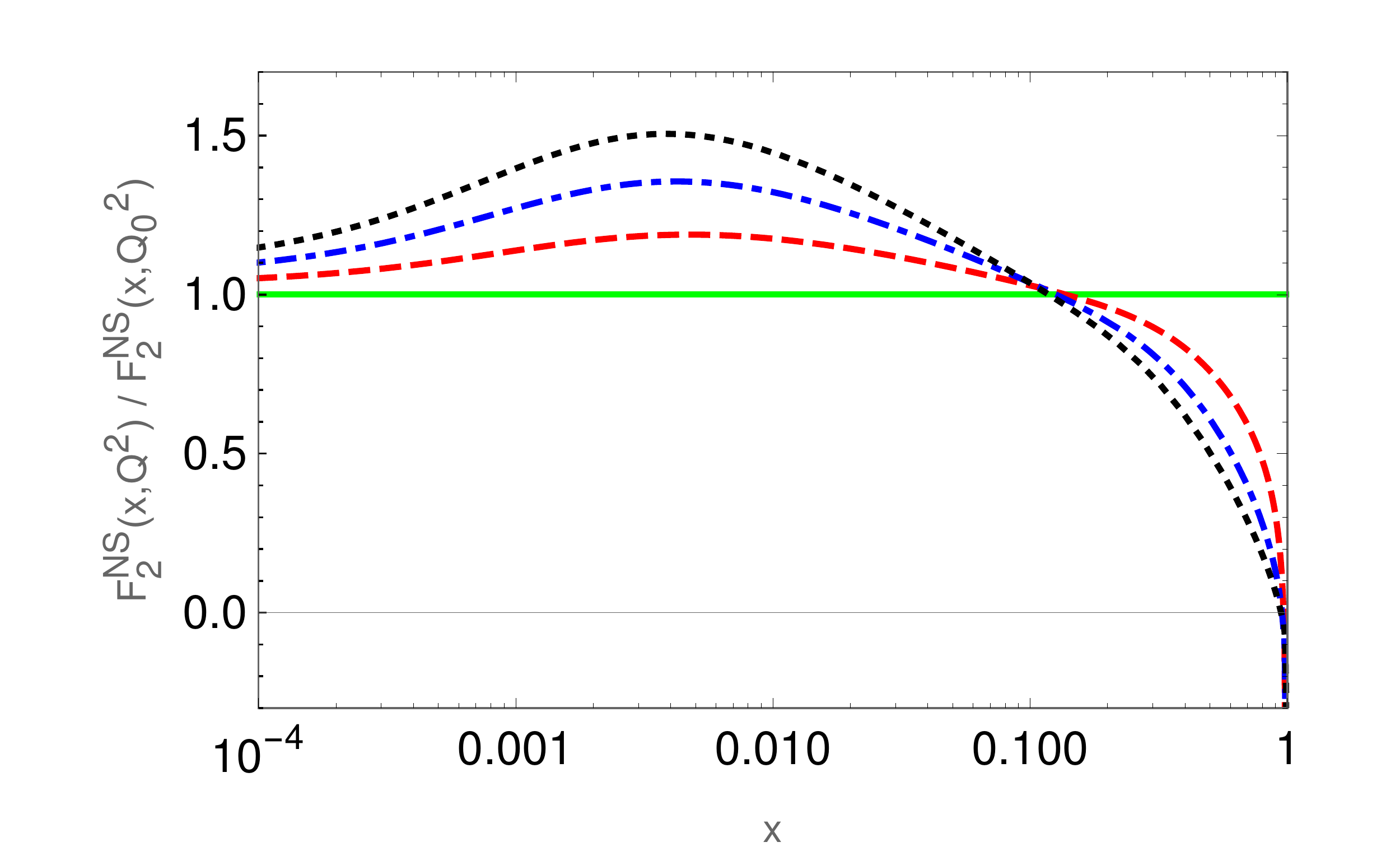}
        \includegraphics[width=0.49\textwidth]{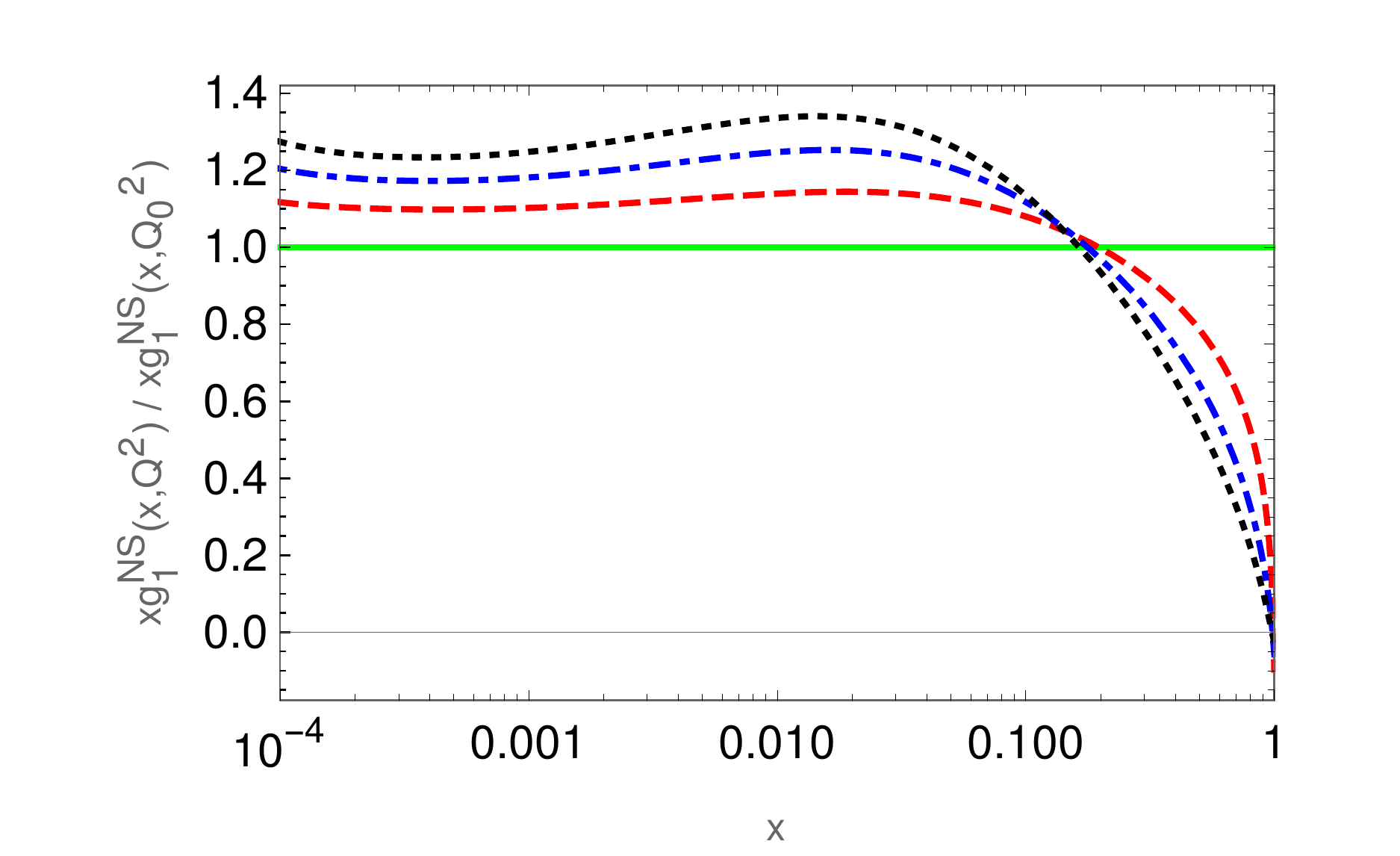}
        \caption{\sf Left:~The relative contribution of $F_2^{\rm NS}$ in the evolution from $Q^2 = 10~\GeV^2$ to $10000~\GeV^2$.
        Right:~The same for the structure function $xg_1^{\rm NS}$.
\label{fig3}}
\end{figure}
%----------------------------------------------------------------------------
\noindent
In Figures~\ref{fig3} we illustrate the relative effect of the scale evolution in $Q^2$ both for 
$F_2^{\rm NS}$ and $x g_1^{\rm NS}$ comparing to the starting scale $Q_0^2$. There is a fixpoint 
at $x_0 \sim 0.1$ with a positive evolution with $Q^2$ at lower $x$ and a negative evolution at larger $x$.
%----------------------------------------------------------------------------
\begin{figure}[H]
        \centering
        \includegraphics[width=0.49\textwidth]{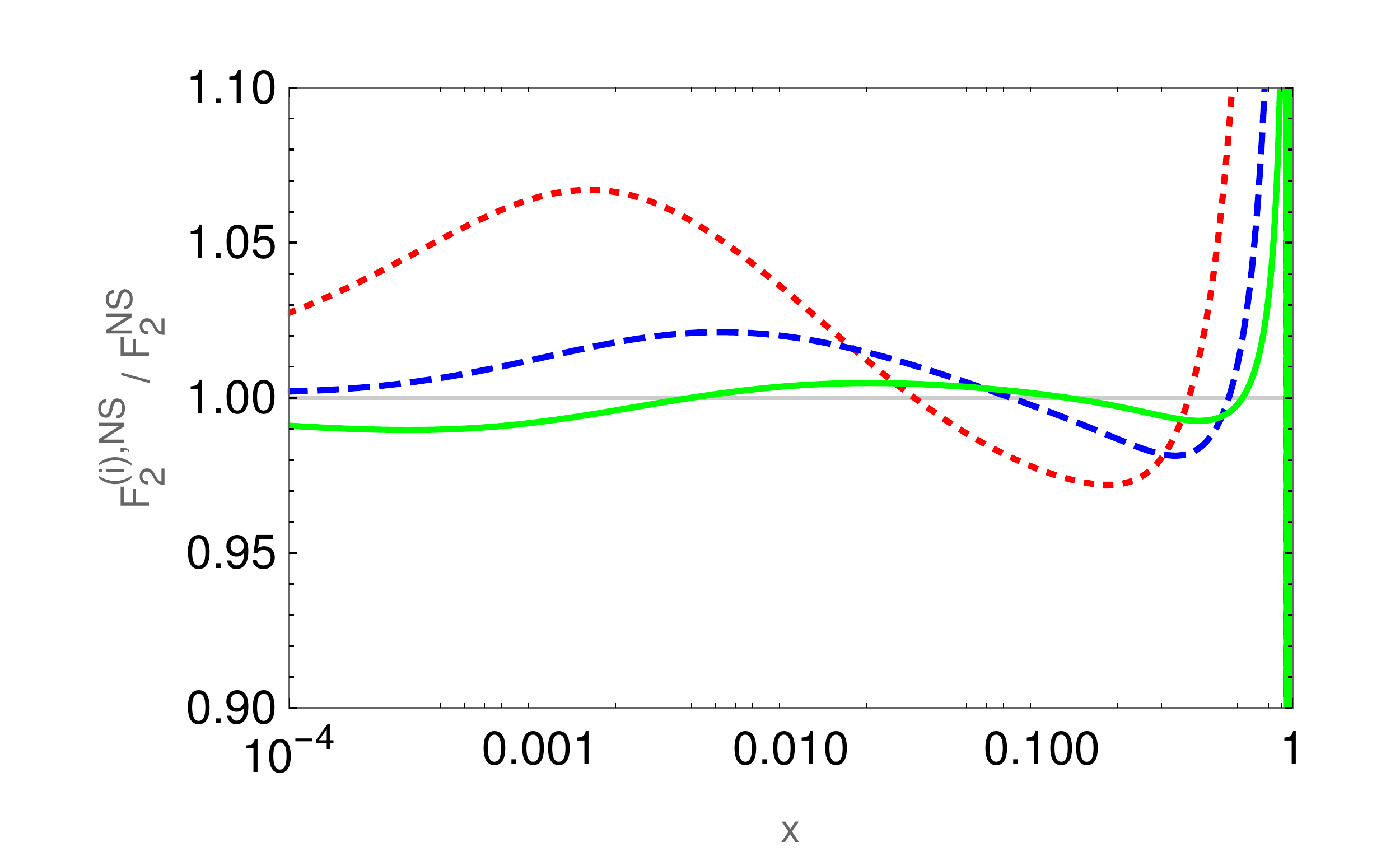}
        \includegraphics[width=0.49\textwidth]{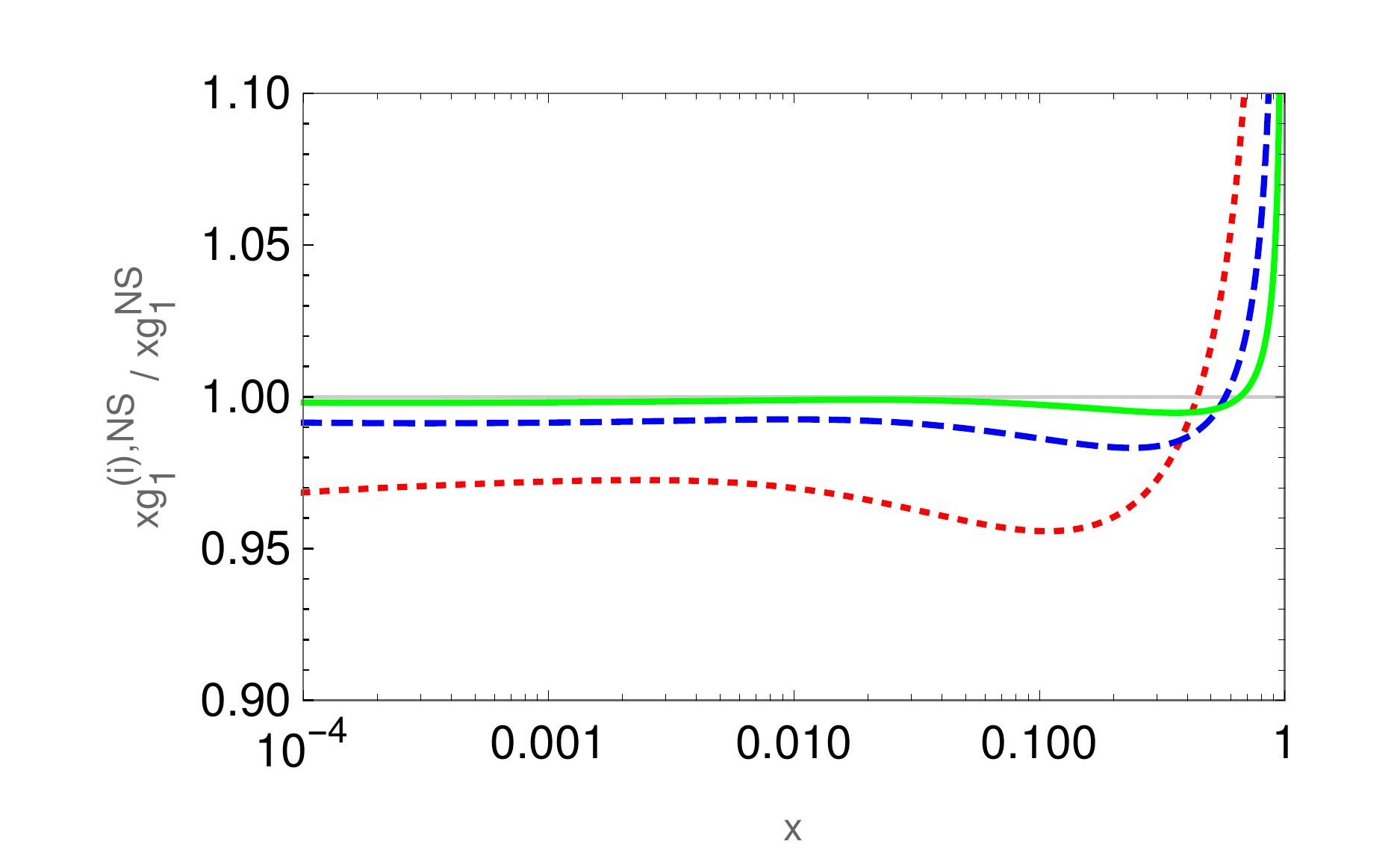}
        \caption{\sf Left:~The relative contributions from LO (dotted lines), NLO (dashed lines) and NNLO (full lines)
to the structure function
$F_2^{\rm NS}$ at N$^3$LO at $Q^2 = 100~\GeV^2$ as an example. Right:~The same for the structure function $xg_1^{\rm NS}$.
\label{fig4}}
\end{figure}
%----------------------------------------------------------------------------
\noindent
In Figures~\ref{fig4} we show the ratio of the results obtained at leading order (LO), 
next-to-leading order (NLO), and next-to-next-to leading order (NNLO) to the N$^3$LO results
at $Q^2 = 100~\GeV^2$. This illustrates the convergence of the perturbative series of the corrections and the 
necessity to include N$^3$LO corrections at an accuracy of the data in the 1\% region.
%----------------------------------------------------------------------------
\begin{figure}[H]
        \centering
        \includegraphics[width=0.49\textwidth]{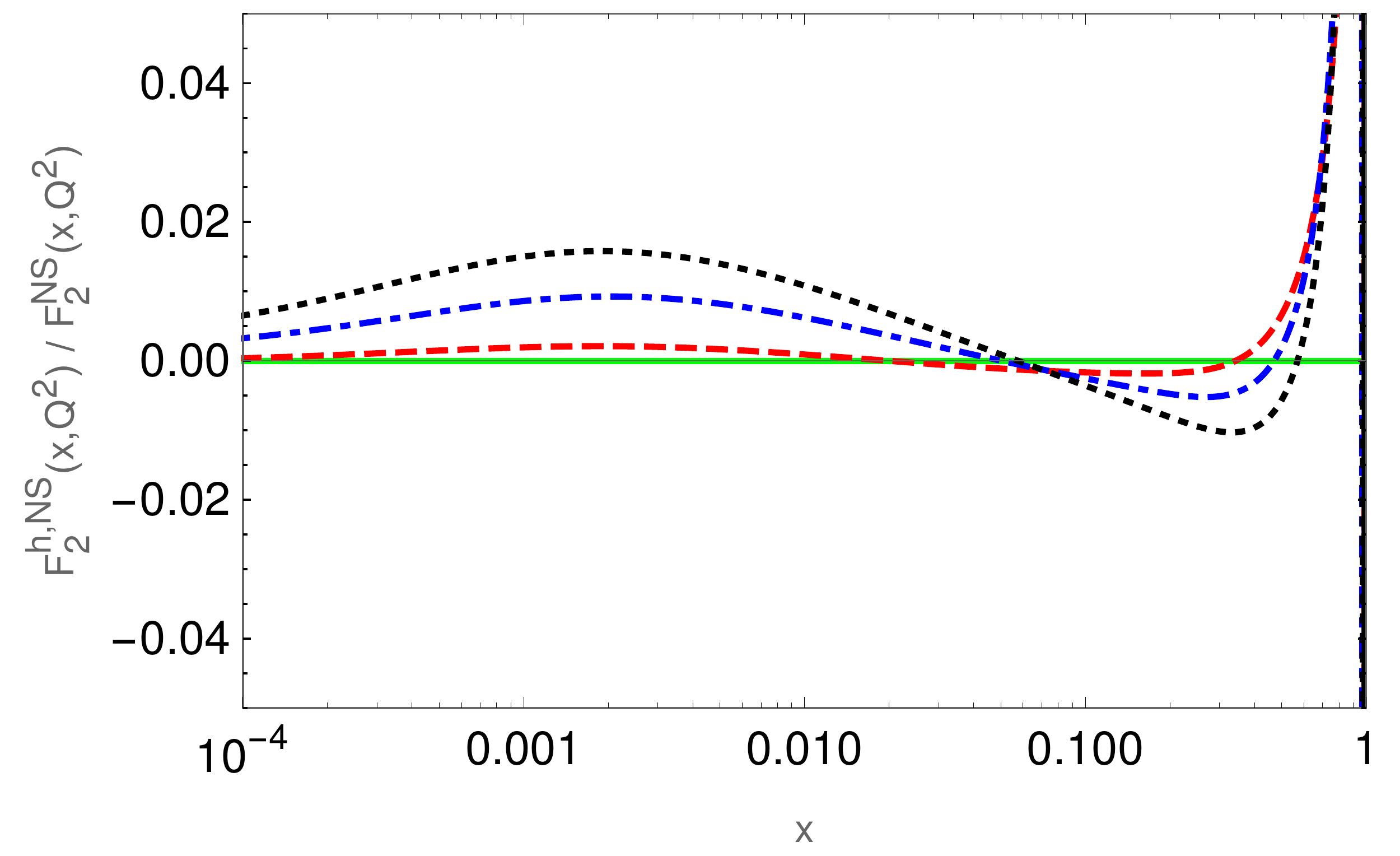}
        \includegraphics[width=0.49\textwidth]{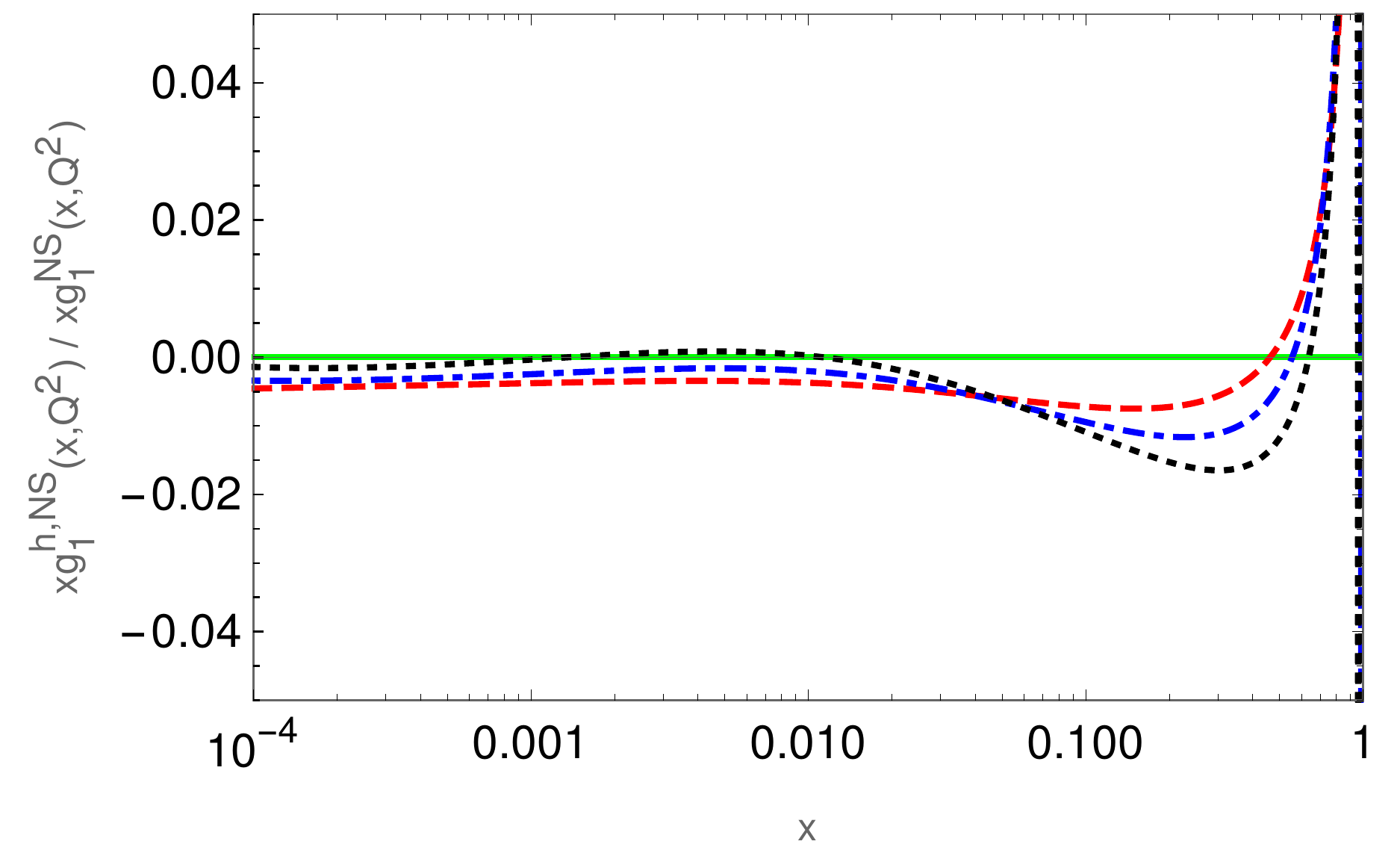}
        \caption{\sf Left:~The relative contribution of the heavy flavor contributions due 
to $c$ and $b$ quarks to the structure function $F_2^{\rm NS}$ at N$^3$LO;
dashed lines: $100~\GeV^2$;
dashed-dotted lines: $1000~\GeV^2$;
dotted lines: $10000~\GeV^2$. Right:~The same
for the structure function $xg_1^{\rm NS}$ at N$^3$LO.} 
\label{fig5}
\end{figure}
%----------------------------------------------------------------------------

In Figures~\ref{fig5} we illustrate the relative size of the heavy flavor parts for the same region in $Q^2$ in the 
unpolarized and polarized cases.  In the important region $x \leq 0.4$ the heavy flavor corrections reach the size of 
$\sim 1\%$. Therefore they are of importance for precision analyses. Larger relative effects are found at larger values of $x$.
There, however, the structure functions and their heavy flavor content drop rapidly.

\begin{figure}[H]
        \centering
        \includegraphics[width=0.49\textwidth]{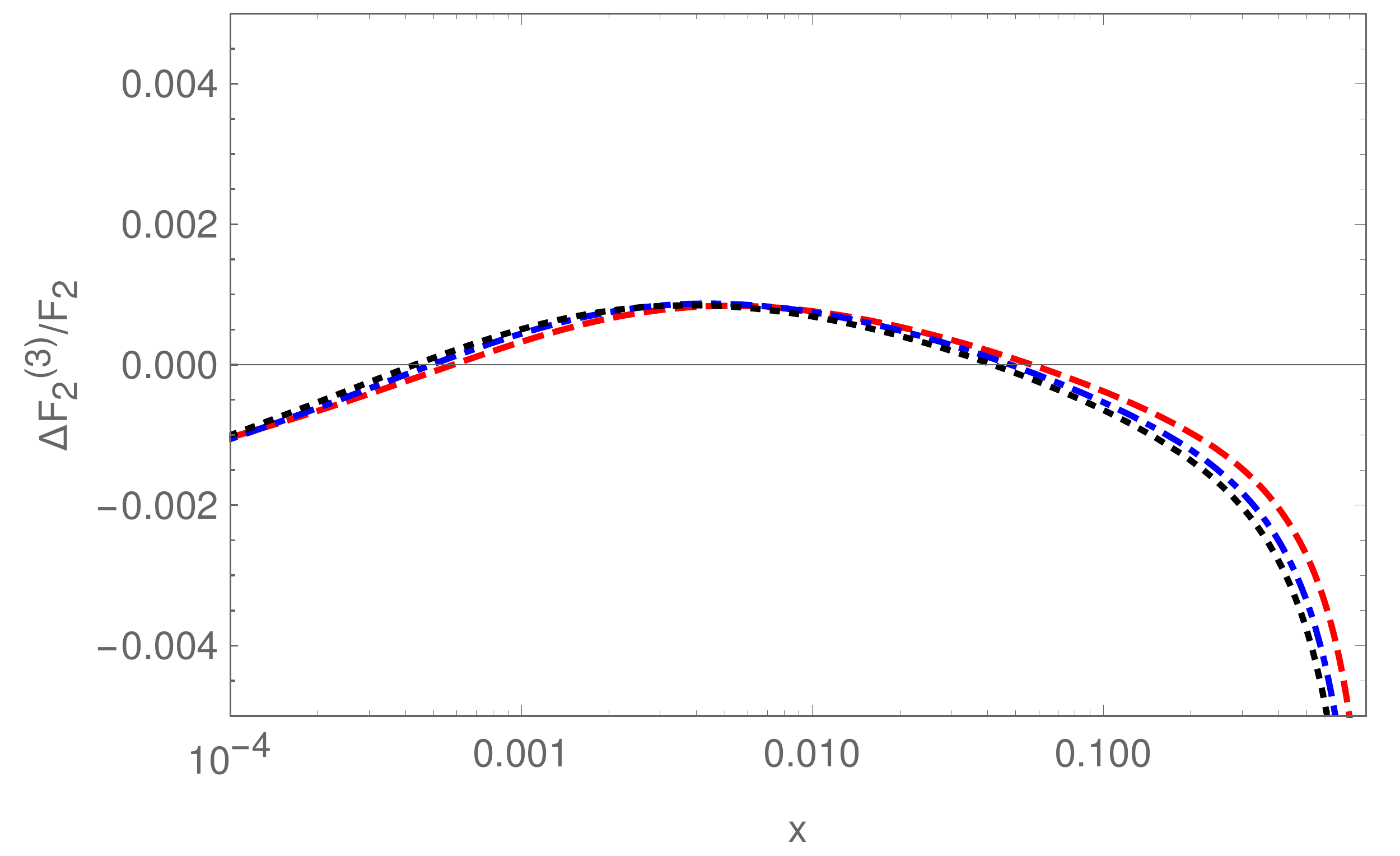}
        \includegraphics[width=0.49\textwidth]{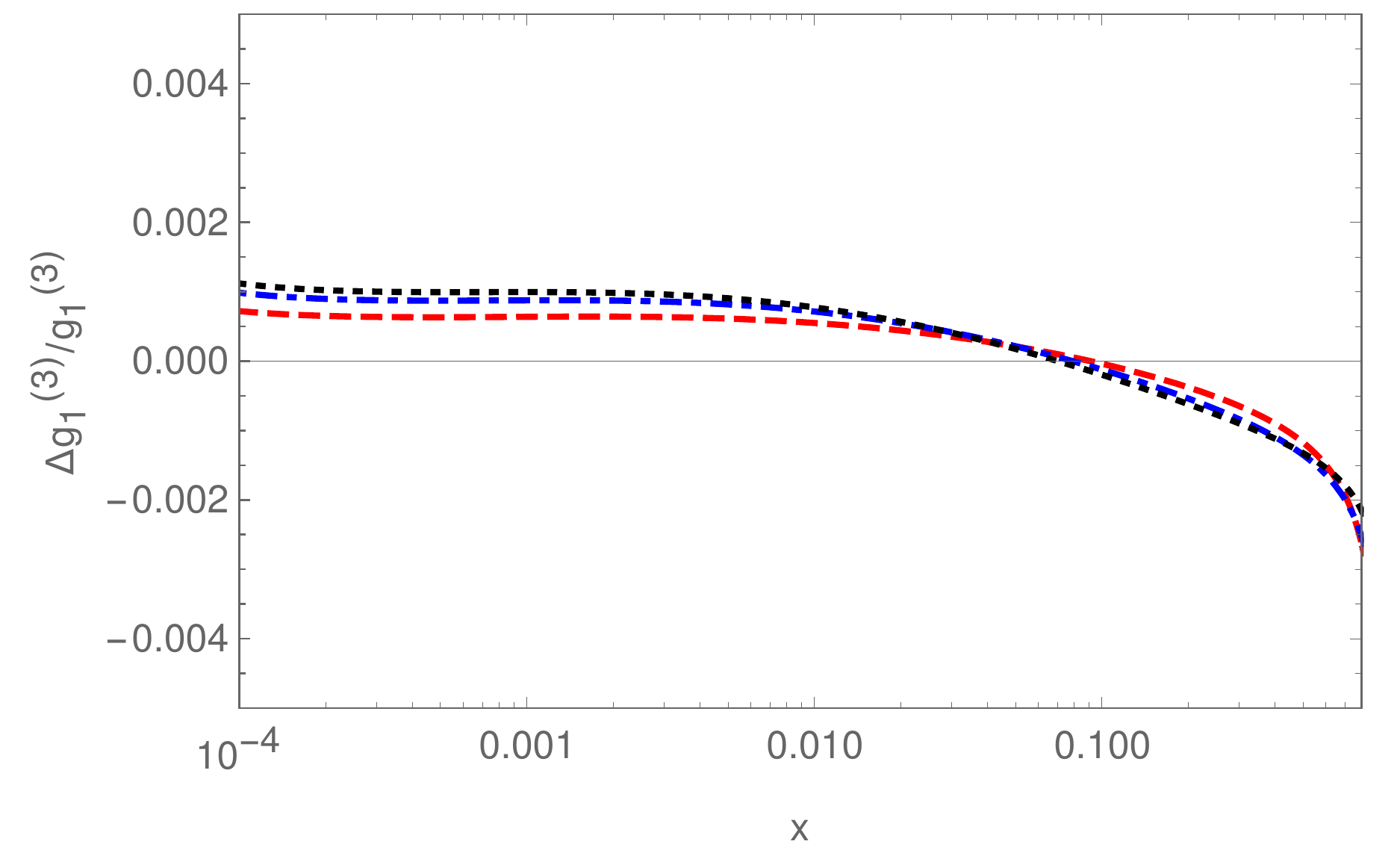}
        \caption{\sf The effect of the variation of $P_{qq}^{3, \pm, \rm NS}$ around the value in Eq.~(\ref{eq:PADE}) by $\pm 
100\%$. Dashed lines: $Q^2 = 100~\GeV^2$, Dash-dotted lines: $Q^2 = 1000~\GeV^2$; Dotted lines: $Q^2 = 10000~\GeV^2$.
Left: $F_2^{\rm NS}$; Right: $xg_1^{\rm NS}$.}
\label{fig6}
\end{figure}

In Figures~\ref{fig6} we illustrate the effect of the half difference if putting $P_{qq}^{3, \pm, \rm NS} = 2 {P_{qq}^{2, \pm, 
\rm NS}}^2/ P_{qq}^{2, \pm, \rm NS}$ and $P_{qq}^{3, \pm, \rm NS} = 0$ 
for both $F_2^{\rm NS}$ and $xg_1^{\rm NS}$. This rescaled correction is in the 
sub--percent range. Moreover, the impact on $\Lambda_{\rm QCD}$ comes from the slope in $Q^2$ which is seen to be rather small.
%----------------------------------------------------------------------------
%----------------------------------------------------------------------------
%-----------------------------------------------------------------------------------------------------------
\section{Conclusions}
\label{sec:6}
%-----------------------------------------------------------------------------------------------------------

\vspace*{1mm}
\noindent
We have calculated the evolution of flavor non--singlet structure functions in the unpolarized and polarized 
case, which become available in high luminosity experiments performed both on proton and deuteron data with
comparable statistics. The evolution has been performed in Mellin $N$ space using our code {\tt QCDEVO}.
Referring to the structure functions directly has the advantage that the input distributions 
are measured. The evolution does only depend on the QCD scale $\Lambda_{\rm QCD}$ and the heavy 
quark masses $m_c$ and $m_b$, the latter of which have been measured very precisely already \cite{PDG}. 
Therefore, this method allows for a one parameter fit of $\Lambda_{\rm QCD}$ only, and provides a method 
which is systematically very stable. The input, after corrections for the deuteron wave function, is given by 
the difference of two structure functions which both can be measured at very high statistics at future facilities
\cite{Boer:2011fh,AbelleiraFernandez:2012cc}. The 
heavy flavor corrections amount to  a contribution of $\sim 1\%$ in the region $x < 0.4$.

Let us finally mention that one may also consider the scheme--invariant singlet 
evolution~\cite{Bardeen:1979uf,Buras:1979yt,Floratos:1981hs,
Furmanski:1981cw,Gluck:1981sc,Grunberg:1982fw,YNDURAIN,Catani:1996sc,Thorne:1997mb,Blumlein:2000wh,Blumlein:2004xs}
in addition to the one in the flavor non--singlet case, provided the corresponding flavor decomposition can be 
carried out. This will usually require more than just the deep--inelastic data, because of the sea--quark 
combinations. A possible way would consist in referring to the Drell--Yan cross section here. Already this part 
complicates the 
analysis. Furthermore, one needs to measure the slope $\partial F_i/\partial \ln(Q^2)$ at the input scale $Q_0^2$ 
together with $F_i(Q_0^2)$ in an uncorrelated way. A second possibility would consist in using the pair $F_2^{\rm S}(Q_0^2), 
F_L^{\rm S}(Q_0^2)$. However, the measurement of the structure function $F_L^{\rm S}$ is much more difficult than 
that of $F_2^{\rm S}$ and in the past not enough statistics has been collected in this case, 
cf.~\cite{Adloff:1996yz,Andreev:2013vha,Chekanov:2009na}, see, however, \cite{LHeC:2020van}.
To perform the necessary flavor decomposition, one needs to perform all these measurements both at proton and 
deuteron targets.
By scheme--invariant evolution the fitting problem of input distributions does not exist and no further optimization by 
neural network techniques is necessary since the input distribution is fully determined by its 
measurement.

\vspace{2ex}
\noindent
{\bf Acknowledgment.}~We would like to thank K.~Chetyrkin, K.~Sch\"onwald, and A.~Vogt for discussions. 
This project has received funding from the European Union's Horizon 2020 research and innovation programme 
under the Marie Sk\l{}odowska-Curie grant agreement No. 764850, SAGEX.

{\footnotesize
%----------------------------------------------------------------------------------------------------------

%----------------------------------------------------------------------------------------------------------
}
\end{document}